# *Assessing Chronic Kidney Disease from Office Visit Records Using Hierarchical Meta-Classification of an Imbalanced Dataset*

Moumita Bhattacharya[1], Claudine Jurkovitz, MD, MPH[2] and Hagit Shatkay, PhD[1,3,4]

[1]Computational Biomedicine Lab, Computer and Information Sciences, University of Delaware, Newark, DE, USA
[2]Value Institute, Christiana Care Health System, Newark, DE, USA
[3]Center for Bioinformatics and Computational Biology, Delaware Biotechnology Inst, University of Delaware, Newark, DE, USA
[4]School of Computing, Queen's University, Kingston, ON, K7L 3N6, Canada
{moumitab, shatkay}@udel.edu

*Abstract — Chronic Kidney Disease (CKD) is an increasingly prevalent condition affecting 13% of the US population. The disease is often a silent condition, making its diagnosis challenging. Identifying CKD stages from standard office visit records can help in early detection of the disease and lead to timely intervention. The dataset we use is highly imbalanced. We propose a hierarchical meta-classification method, aiming to stratify CKD by severity levels, employing simple quantitative non-text features gathered from office visit records, while addressing data imbalance. Our method effectively stratifies CKD severity levels obtaining high average sensitivity, precision and F-measure (~93%). We also conduct experiments in which the dimensionality of the data is significantly reduced to include only the most salient features. Our results show that the good performance of our system is retained even when using the reduced feature sets, as well as under much reduced training sets, indicating that our method is stable and generalizable.*

## I. INTRODUCTION

Chronic kidney disease (CKD) is an increasingly prevalent condition affecting *13%* of the US population [1]. It is defined as kidney damage that persists for more than three months, and is typically stratified into five stages, *1-5*, indicating increasing order of severity [2]. CKD severity is quantified by *estimated glomerular filtration rate* (*eGFR*) [3], an indicator of the level of kidney function. Stage *1* is defined by kidney damage (protein or blood in the urines) while eGFR is normal (eGFR $\geq 90$ *ml/min/1.73m²*); stage *2* by kidney damage and mildly decreased eGFR (eGFR *60-<90*); stage *3* as eGFR *30-<60*; stage *4* as eGFR *15-<30* and stage *5* as eGFR *<15*. The eGFR measure is estimated from *serum creatinine lab tests*, *race*, *sex* and *age*. CKD patients, especially those at stages *4* and *5*, are at high risk of end stage renal disease (ESRD) or death if their condition is not diagnosed [1]. However, because CKD, even in its advanced stages, is often asymptomatic, serum creatinine lab tests are often not ordered, resulting in the condition remaining undiagnosed.

According to a report by the Kidney Early Evaluation Program (KEEP) [1], among *122,502* program participants enrolled at the time of the study, fewer than *30%* of patients at stages *4* and *5* had ever seen a nephrologist. However, *95%* of these patients had visited a general practitioner during the year preceding the study, for a condition other than CKD. Hence, a risk stratification model that separates CKD patients into severity stages based on information gathered during office visits can alert physicians about the advanced stages (stages *4-5*) of the condition, and prompt them to urgently order the appropriate lab tests confirming the diagnosis.

According to treatment guidelines, patients at CKD stage *4* should be referred to a nephrologist and be prepared for renal replacement therapy [3]. The latter typically includes hemodialysis, which patients often begin with only a short notice, and thus without the benefit of a functional arteriovenous fistula, which needs to be surgically created several months prior to hemodialysis initiation. Recent data from the United States Renal Disease System (USRDS) show that only *17%* of patients initiating hemodialysis do so with a functional arteriovenous fistula [4]. Finally, given the increasing focus on population health management, identifying patients at high risk for end stage renal disease from *Electronic Health Records* (*EHRs*), may be helpful for case-managers that are often responsible for the health of thousands of individuals. Hence, in this study, we aim to stratify CKD patients into severity stages (CKD stages *3-5*), based on clinical information gathered during office visits.

In collaboration with physicians from Christiana Care Health System, the largest health-system in Delaware, we analyze a dataset gathered over a period of nine years from *13,111* patients. It comprises *120,739* records obtained during patients' visits to multiple primary care and specialty practices across Delaware, which are part of the information stored in EHRs. Patient records were selected for inclusion in the dataset if at any time during follow-up, there was an indication of a decline in kidney function, determined by a lower than normal eGFR value (*<60*), indicative of CKD at stage *3* or higher. The resulting dataset includes all records of patients associated with stages *3*, *4* and *5*. Each record in the dataset comprises *495* simple quantitative non-text attributes summarizing the patient's *demographics, vital signs, medications* and *diagnosed conditions*. We use the values of these attributes as features to represent each patient's visit record. Notably, unlike physicians' notes, these *495* non-text attributes are available for the vast majority of patients, and their semantics is unambiguous and readily interpretable. The dataset is described in Section II.

While EHRs provide valuable patient health information for patient risk-stratification and disease prediction, one of the major challenges in using them arises from *data imbalance*. That is, only a small proportion of patients suffer from the more severe conditions, while most patients suffer only a milder manifestation of the disease. Specifically, in our dataset, the number of records associated with stage *3* is *10* times larger than the number of records associated with stage *4*, and *23* times larger than that of stage *5*. Imbalanced datasets, whose class distribution is skewed, form a common challenge in data mining applications, such as fraud or disease detection, where the class of interest is heavily underrepresented compared to the other classes. As noted later in this section, learning classifiers using off-the-shelf packages from such an imbalanced dataset typically leads to poor performance.

We thus propose and develop a supervised machine learning method that addresses data imbalance, while aiming to stratify CKD patients already identified as advanced, into severity stages (stages *3-5*), using information gathered from standard office visit records. Taking advantage of such standard records supports a generalizable approach, applicable to most patients who regularly see a healthcare provider.

Additionally, to extract from the records the most informative features indicative of decline in kidney function, we conduct feature selection using *Least Absolute Shrinkage and Selection Operator* (*LASSO*) regression [5], a regularized linear regression method that assigns a weight of *0* to less informative features, ensuring that only relevant features are taken into account in the classification.

Several recent studies proposed approaches for patient risk-stratification and disease prediction using machine learning methods [6-10]. For instance, Mani et al. [6], aimed to predict diabetes risk based on lab results, diagnosed conditions and medications, using six simple classification methods, including naïve Bayes and decision trees. Similarly, Ogunyemi et al. [7] added insurance information and vital signs, while aiming to predict risk of diabetic retinopathy using an ensemble classifier, based on decision tree learners. The authors use random under-sampling of the majority class to address data imbalance. Teixeira et al. [8] also used similar information while including narrative text aiming to predict hypertension using the random forest classifier. Unlike our study, none of the above is based solely on simple quantitative attributes in office visit records; they instead use lab test results, insurance information and narrative text, in addition to office visit records. Moreover, all three studies were based on only very small datasets (< *2,300* records), while ours is an order of magnitude larger. Last, while aiming to conduct prediction, the training set in these studies was not limited to temporally early records while restricting the test set to later records, as we do here. As such, these studies do not actually demonstrate predictive power.

In a significantly larger study (n≈*35,000*), Huang et al. [9] aimed to predict depression severity and patient's response to treatment, using free-text clinical reports in addition to office visit records. Like we do here, they too employ the *LASSO* regression for feature selection and prediction. However, they do not directly address class imbalance that is inherent in the dataset, as we do here, and their reported performance stands at *50%* sensitivity at the time of diagnosis and *25%* sensitivity *12* months prior to diagnosis.

To the best of our knowledge, ours is the first study aiming to identify disease stage exclusively using simple quantitative attributes obtained from office visit records, while also directly addressing the class imbalance among different severity levels.

Data imbalance is often handled in machine learning via sampling strategies [11-17]. Typically, these involve either *under-sampling* – reducing the size of the majority class by removing instances from the training set, or *over-sampling* – increasing the impact of the minority class, by sampling from it with repetition. Several previous studies have proposed variants of under/over-sampling approaches to address class imbalance. Examples include one-sided selection [11] and *synthetic minority oversampling technique* (*SMOTE*) [12].

Notably, most of these are concerned with binary classification, while ours is a multiclass task, as we are looking to assign one of three possible stages to each record (possibly more in future studies).

Two approaches commonly used to transform multi-class classification into multiple binary-classification tasks are *one-against-all* (*OAA*) and *one-against-one* (*OAO*) [13]. Tan et al. [14] use both these schemes in the context of protein fold classification, and subsequently build rule-based learners to improve coverage of the minority class. Zhao et al. [15] use *OAA* to handle multiclass classification, while employing under-sampling and *SMOTE* techniques on imbalanced data, for protein classification. Another approach to address class imbalance is using cost sensitive ensemble methods [16].

Before developing our own approach, we have applied versions of the above methods, specifically, random over- and under-sampling, and over-sampling using *SMOTE* and *ADASYN*. None has improved on the results obtained by simple classifiers that do not account for class imbalance (to which we refer as *baseline classifiers*), such as simple random forest. Thus, as mentioned earlier, we developed a supervised machine learning method, *hierarchical meta-classification*, that aims to separate CKD stages *3*, *4* and *5*, while addressing data imbalance. The method frames the multiclass classification task as a sequence of two subtasks. The first is binary classification, separating records associated with stage *3* from those associated with a combined class consisting of stages *4* and *5*, using meta-classification. *Meta-classification* combines results obtained from multiple simple classifiers (base-classifiers) into a single classification decision [18]. The second subtask separates the records assigned to the combined stages *4* and *5* class into the individual classes.

To train the hierarchical meta-classifier, we take advantage of the earlier visit records, gathered between the years *2007-2014*, while testing is done based on later records, gathered during *2015*. We also trained the hierarchical meta-classifier on a dataset represented by only the informative attributes identified through feature selection, and compared the performance to that obtained when the complete set of features was used. We evaluate the performance of our methods using standard metrics, namely, *overall accuracy*, as well as *specificity, sensitivity, precision* and *F-measure* [5]. The results demonstrate that our hierarchical meta-classifier trained on a dataset represented via the complete set of features, improves upon multiple baselines, showing a performance level of about *93%* according to almost all evaluation measures, along with an average specificity of *77%*. The good performance of our method indicates that simple quantitative attributes from office visit records can form the basis for CKD severity detection and stratification. Our results also indicate that our method retains its high performance level even when trained on a dataset represented via a reduced feature set. Furthermore, to assess the medical validity of our results we verify that the features identified as pertinent by the feature selection process are indeed indicative of CKD severity according to the medical literature.

To further demonstrate the predictive ability of our method, we progressively reduced the training set size by truncating the early years of patient history, one year at a time, yielding *8* training sets. The test set was fixed to be the records collected in *2015*, which ensures that the training set always comprises temporally earlier records than the test set. We show that classification remains as effective when the number of years

(and of visit records) included in the patient history and used by the classifier is reduced, illustrating the stability and generalizability of our hierarchical meta-classifier.

## II. DATASET

Our dataset comprises *120,739* records obtained during patients' visits to multiple primary care and specialty practices across Delaware; these records form part of the information stored in EHRs. As mentioned earlier, records were included in the dataset if they indicated a decline in kidney function, determined by a lower than normal eGFR value (< *60 mL/min/1.73m$^2$*), indicating CKD of stage *3* or higher.

We removed from the dataset *27,521* records that were missing essential values, leaving a set of *93,218* complete records, where *60%* are associated with female patients and *40%* with male. The average number of visits per patient, over the *9* year period, is *9*; the patients' mean age is *70 ($\sigma$ =12.4)*, and the range is *18-107*, where *70%* of the patients are *58-82*. Table 1 shows the four categories of features comprising the dataset, along with the number of actual features per category. Information pertaining to the features listed in Table 1 is routinely collected and stored in the EHR during each visit to the general practitioner, making our approach broadly applicable to office visit records beyond the specific disease and dataset analyzed here.

Given our aim of stratifying CKD by severity, we have removed five of the *495* features that are directly reflective of CKD from our feature set. The five features denote five diagnosed CKD conditions, namely: CKD stage *2*, CKD stage *3*, CKD stage *4*, End Stage Renal Disease and Chronic Renal Failure. We thus use a total of *490* features for representing the patients' records in our dataset. Specifically, each record, $r^k$ ($1 \leq k \leq 93,218$), is represented as a *490*-dimensional vector, $V^k = < v_1^k, ..., v_{490}^k >$, where each dimension corresponds to one of the *490* features.

As noted in Section I, our dataset is highly imbalanced, as is often the case in a biomedical setting, where the outcome of interest, in our case, stages *4* and *5*, is rare, and thus underrepresented. In our study, the ratio among the number of records associated with each of the stages *3*, *4* and *5* is *23:2:1* respectively. Recall that we progressively reduced the training set size by pruning the early years of patient history, one year at time, yielding *8* training sets. Table 2 shows the number of records per class, for each of the *8* training sets. The table also shows the number of records per stage collected in *2015* and used as the test set.

## III. METHODS

We next describe the hierarchical meta-classifier we have developed, including the feature selection method and the baseline and simple meta-classifier used for comparison.

**Feature Selection**: Building classifiers from a large feature set can lead to over-fitting. We thus conduct feature selection to determine the most relevant attributes for CKD stratification using *LASSO* regression [5], a modified form of least squares regression that penalizes model complexity via a regularization parameter by assigning zero weights to less informative features. For training the regression model, we use eGFR values instead of CKD stages as the response variable, since eGFR is a continuous numeric variable. The salient features predictive of eGFR are also informative of CKD severity-levels, since the latter are calculated from eGFR values. We use the *Python scikit-learn* package [19] for applying *LASSO* regression.

**Baseline Classifiers:** As a baseline for comparison, we use four standard methods, namely, logistic regression, naïve Bayes, decision tree and random forest, employing the one-against-all strategy to assign a CKD stage to each office visit record. We use the *Python scikit-learn* implementation for training the four classifiers [19].

TABLE 1. FOUR CATEGORIES OF FEATURES COMPRISING OUR DATASET. THE LEFTMOST COLUMN SHOWS THE CATEGORIES, WHILE THE MIDDLE COLUMN SHOWS THE NUMBER OF FEATURES PER CATEGORY. THE RIGHTMOST COLUMN SHOWS EXAMPLES OF FEATURES ASSOCIATED WITH EACH CATEGORY.

| Category | Number of Features | Examples |
|---|---|---|
| ***Demographics*** | 4 | *Gender*; *Age*; *Ethnicity*; *Race* |
| ***Vital Signs*** | 4 | *Heart Rate*; *Systolic and Diastolic Blood Pressure, Body Mass Index* |
| ***Diagnosed Conditions*** | 454 | *Benign essential hypertension*; *Type 2 diabetes mellitus*; *Obesity* |
| ***Medications*** | 33 | *ACE Inhibitors*; *Alpha Beta Blocker*; *Insulin* |

TABLE 2. NUMBER OF RECORDS IN EACH OF THE CLASSES WITHIN THE 8 TRAINING SETS OBTAINED BY PROGRESSIVELY TRUNCATING THE EARLY YEARS OF PATIENT HISTORY INCLUDED IN THE TRAINING DATA. THE RESPECTIVE YEAR RANGE IS SHOWN IN THE SECOND ROW. THE LEFTMOST COLUMN SHOWS CKD STAGES. THE RIGHTMOST COLUMN SHOWS THE NUMBER OF RECORDS PER STAGE, COLLECTED IN 2015 AND USED AS THE TEST SET. EACH OF THE OTHER COLUMNS SHOWS THE NUMBER OF RECORDS PER STAGE FOR THE CORRESPONDING TRAINING SET, COLLECTED DURING THE PERIOD INDICATED IN THE SECOND ROW.

| CKD Stages | TRAINING SET DISTRIBUTION | | | | | | | | TEST SET |
|---|---|---|---|---|---|---|---|---|---|
| | **2007-2014** | **2008-2014** | **2009 - 2014** | **2010-2014** | **2011-2014** | **2012-2014** | **2013-2014** | **2014** | **2015** |
| **Stage 3** | 73,425 | 72,808 | 70,127 | 65,326 | 57,863 | 46,881 | 33,072 | 17,273 | 8,419 |
| **Stage 4** | 6,976 | 6,903 | 6,579 | 6,060 | 5,385 | 4,439 | 3,101 | 1,624 | 782 |
| **Stage 5** | 3,241 | 3,184 | 3,052 | 2,821 | 2,515 | 2,068 | 1,471 | 767 | 375 |
| ***Total*** | *83,642* | *82,895* | *79,758* | *74,207* | *65,763* | *53,388* | *37,644* | *19,664* | *9,576* |

*Meta-Classification*: To separate CKD stages *3–5*, while addressing data imbalance, we employ meta-classification, which aims to assemble results obtained from multiple classifiers into a single classification outcome. The approach comprises two sub-tasks: First, a set of *M* simple classifiers, $\{C_1,...,C_M\}$ are trained, and applied to each visit record, $r^k$, where the latter is represented as a *490*-dimensional vector, $<v_1^k,...,v_{490}^k>$ as described above. We refer to the simple classifiers as base-classifiers. Each of these classifiers assigns a label $C_j^k$ ($C_j^k \in \{3,4,5\}$) to the vector $V^k$. Second, the class labels assigned by the *M* simple classifiers are used to re-represent the visit record $r^k$ as an *M*-dimensional vector $<C_1^k,...C_M^k>$. This representation is then used to train a meta-classifier that assigns the final class label to each record [18].

To train the base-classifiers, we produce balanced training sets by first partitioning the data stemming from the over-represented classes into smaller subsets. Specifically, as the ratio among the number of records associated with each of the stages *3*, *4* and *5* is *23:2:1* respectively, we partition the stage *3* set into *23* equal subsets and the stage *4* set into two equal subsets. Each subset contains the same number of records as that included in the stage *5* set. Next, we combine each of the stage *3* subsets with the stage *5* set and with one of the two stage *4* subsets that we select at random, thus forming a total of *23* datasets, each having a uniform distribution across the three CKD stages. Fig. 1 illustrates the data partitioning scheme.

To choose the base classifier, from among four commonly used simple classifiers, namely logistic regression, naïve Bayes, decision tree and random forest, we conducted four sets of experiments, in each we employed one of these methods as a base classifier. We trained each of the four classifier types on the *23* sets, thus generating *23* base-classifiers per type. Using each set of *23* base-classifiers, we trained a meta-classifier in which the training set was re-represented as *23*-dimensional vectors (*M=23*), where the value along the $i^{th}$ dimension consists of the label obtained from the $i^{th}$ base-classifier when applied to the original record representation. In all four sets of experiments the meta-classifier used is a naïve Bayes classifier, as it proved to perform best, and has proven effective by others as well [20].

*Hierarchical Meta-Classification*: As CKD stage *3* (eGFR range of *30–60*) is characterized by moderately reduced kidney function, while stages *4* and *5* (eGFR < *30*) are characterized by severely reduced kidney function [1], records associated with patients at stages *4* and *5* are likely to be more similar to one another than to records of patients at stage *3*. Thus, the simple meta-classifier introduced above often fails to separate between stages *4* and *5*, while also misclassifying stage *3* records. Therefore, we introduce hierarchical meta-classification, which first separates records associated with stages *4* and *5* from those associated with stage *3*, and then further separates the combined class consisting of stages *4* and *5* records into the two individual subclasses. To separate the stage *3* class from the combined stage *4* and *5* class we use the meta-classification scheme described above. We then employ multiple simple classifiers (including random forest and naïve Bayes), to separate the combined stage *4* and *5* set into the two respective stages. The simple random forest classifier is most effective in separating stage *4* from stage *5* patients.

Fig. 2 summarizes the hierarchical meta-classification approach, where the top dashed-square corresponds to the *coarse high-level classification*, separating stage *3* from the combined stages *4* and *5* records, while the bottom dashed-square depicts the *refinement step*, separating the combined class into the refined stage *4* and stage *5* classes.

To train the base-classifiers for the coarse classification step (distinguishing stage *3* from the combined two other stages), we first partition the data using a similar scheme to that shown in Fig. 1, albeit partitioning the records associated with stage *3* into only *7* parts (rather than *23*) as there are only *7* times more records associated with stage *3* than with the combined set of stages *4* and *5*. We next train the *7* respective base-classifiers, re-represent the datasets as described when discussing meta-classification above, and train a naïve Bayes meta-classifier using the re-represented data. The resulting classifier aims to separate stage *3* records from records that are stage *4* or *5* (see Fig. 2A).

To train the random forest classifier for the refinement step (separating the combined class into individual stage *4* and stage *5* records, see Fig. 2B), we use the set of records from the training data that are associated with stages *4* and *5*.

To test the meta-classifiers, each record in the test set is classified by each of the base-classifiers, and the labels obtained from each of these base classifiers are used to form a feature vector, which becomes the input to the meta-classifier. The meta-classifier is applied to each newly represented vector, thus separating stage *3* records from records of stages *4* or *5*. Records classified into the combined stage *4* and stage *5* class in the coarse classification step, are further categorized by the simple random forest classifier in the refinement step, and assigned to either the stage *4* or stage *5* individual class.

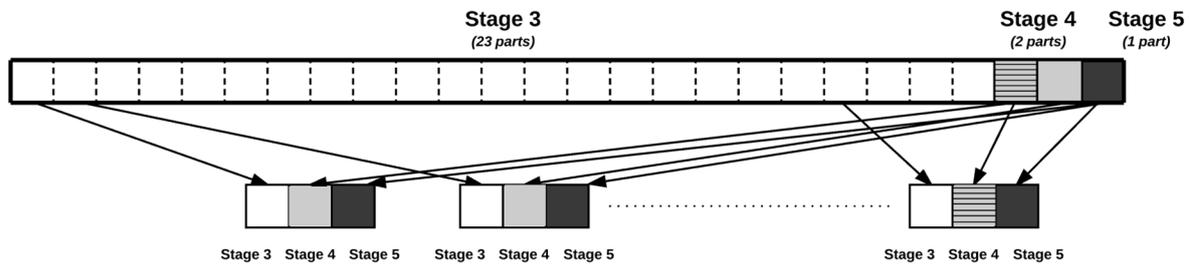

Figure 1. Data partitioning scheme for meta-classification. The set of stage *3* records is partitioned into *23* subsets (white squares in the figure). That of stage *4* is partitioned into two subsets (grey squares and lined squares), where each subset contains the same number of records as that included in the stage *5* set (black squares). We combine each of the stage *3* subsets with the stage *5* set and with one of the two stage *4* subsets that we select at random, thus forming a total of *23* datasets, each having a uniform distribution across the three stages.

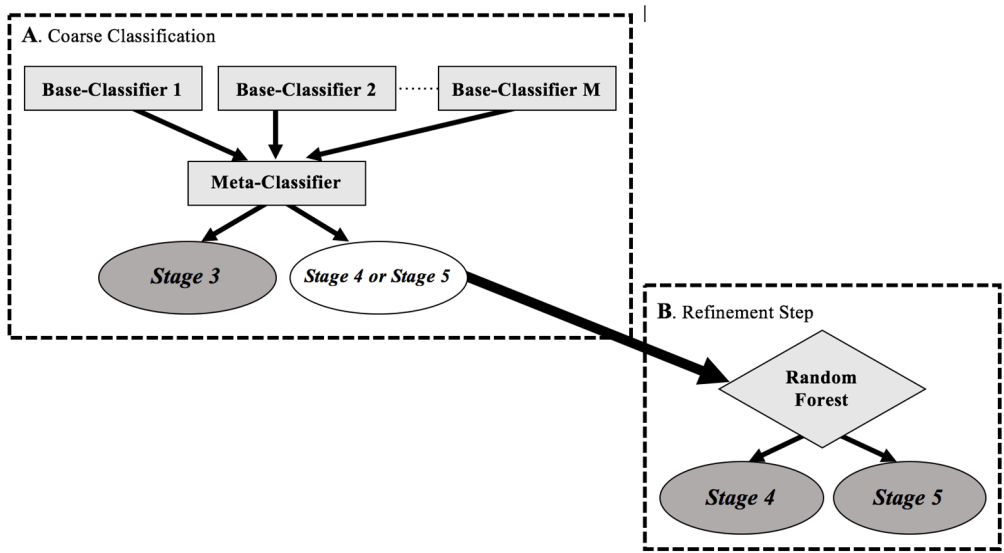

Figure 2. Hierarchical meta-classification for multiclass classification. A. *Coarse classification*: The dashed rectangle represents the meta-classification scheme used for separating stage *3* records from those associated with either stage *4* or stage *5*. The white oval represents the set of records assigned to the combined class by the meta-classifier, while the shaded ovals represent the final classes corresponding to the individual three stages. B. *Refinement step*: The combined set of stage *4* and *5* records is separated into individual respective constituent subsets. The rhombus represents a simple random forest classifier.

## IV. EXPERIMENTS

As a baseline, we first trained and tested simple naïve Bayes, logistic regression, decision tree, and random forest classifiers using the complete set of *490* features to represent each record. Next, to address the data imbalance, we applied both meta-classification and hierarchical meta-classification to the records. For each of the four methods, we used records from the first *8* years (*2007- 2014*) for training, and the ninth year (*2015*) for testing. Notably, we did not use cross-validation for training and testing our methods, since we limited the training set to temporally early records (from the first *8* years, *2007-2014*) while restricting the test set to later records (from the ninth year, *2015*). To ensure stability of the results, we employ multiple random splits to partition the training set stemming from the over-represented class into smaller subsets for training the base-classifiers in the coarse classification stage (see Fig. 2B) of the hierarchical meta-classifier, while keeping the test set fixed to all records from *2015*.

After each classification step, we evaluated the performance using standard measures, namely, overall accuracy, sensitivity, specificity and F-measure. We compared the performance attained from the four classifiers as baseline and as a component of the simple and the hierarchical meta-classifiers, to assess their respective efficacy in separating CKD stages. As noted in Section I, we also applied other methods such as random over- and under-sampling and *SMOTE* to address data imbalance and found the performance of hierarchical meta-classification to be the best. Hence, we focus here only on the results obtained using hierarchical meta-classification.

We next employed feature selection using *LASSO*, to identify the features most indicative of CKD stages. Once the features were selected, we represented patients using the reduced feature set, and repeated the classification experiments to assess the impact of the reduced number of features on performance. We compared the results obtained from the classifiers trained on the dataset represented through the complete feature set with those obtained from the classifiers trained using the reduced representation.

Last, we examined the performance of our method when the number of years (and of visit records) included in the patient history and used by the classifier, is reduced. To do so, we repeated the experiments using the hierarchical meta-classification while progressively truncating the early years of patient history included in the training data, one year at a time, yielding *8* sets of training data. As in previous experiments, we kept the test set fixed to records collected in *2015*. The first set included records gathered between the years of *2007* and *2014*, yielding a training set containing *83,642* records, while the eighth set included data collected in *2014* alone, yielding a training set of *19,664* records. We trained the hierarchical meta-classifier over each of the training sets, using both a record representation employing the complete set of *490* features, and one employing only the reduced feature set. This set of experiments assessed the generalizability of our approach, i.e. its ability to assign the correct severity level based only on the most recent history of the patient.

Recall that to ensure the stability of our model, we repeated each experiment *20* times, using different splits to partition the set of records associated with the over-represented class.

## V. RESULTS

The hierarchical meta-classifier, the baseline classifiers and the simple, non-hierarchical meta-classifier perform significantly better when decision tree or random forest classifiers are used – either alone as baseline classifiers, or in an assemblage of base-classifiers within a meta-classifier – as compared to logistic regression or naïve Bayes. We thus report here only results obtained using decision tree and random forest, as standalone baseline classifiers and as components within meta-classification. Fig. 3 shows the average specificity, sensitivity and F-measure, obtained by the hierarchical meta-classification scheme, compared to those obtained by each of the baseline classifiers and by the simple, non-hierarchical meta-classifier for decision tree and random forest methods. The average is calculated as a weighted-average, taking into account the class-size associated with each of the three stages. Fig. 4 shows the sensitivity and F-measure per-class, attained by the random forest baseline classifier and by the random forest hierarchical meta-classifier.

Our feature selection process identified *119* of the *490* features as most informative of CKD severity-levels. Table 3 lists *18* of the *119* salient features (due to limited space), across the four categories of features comprising the dataset. Table 4 shows a comparison between the performance of the hierarchical meta-classifier trained on the reduced-dimensionality dataset, represented via the selected *119* features, against that of the classifier trained on the dataset represented via the complete feature set.

The above results were obtained from experiments conducted on records dated *2007-14* used for training, while records from *2015* were used for testing. Fig. 5 shows the *True Positive Rates* (*TPR*) and *False Negative Rates* (*FNR*) per-class for the *8* training sets (see Table 2) that were obtained by progressively truncating the early years of patient history in the training data, one year at time. We used each of the *8* sets to train the random forest hierarchical meta-classifier, using the complete feature set. The average accuracy, specificity, sensitivity, precision and F-measure are all about *0.93* (*std < 0.03*) for all *8* sets.

As described in Section IV, we repeated all experiments *20* times using multiple random splits. We obtained similar results in all runs (*std << 0.03*).

VI. DISCUSSION

The hierarchical meta-classifier that employs random forest base-classifiers (denoted *RF-Hier-MC* in Fig. 3) outperforms all other classifiers according to all measures except for specificity (*0.77*), which is lower than that of the simple, non-hierarchical random-forest-based meta-classifier (*0.85*, classifier denoted *RF-MC* in Fig. 3). Accuracy and precision, both *0.93*, are not shown in the figure due to limited space. We also note that the performance of the baseline random forest classifier (denoted *RF-Baseline*) is similar to that of the random forest hierarchical meta-classifier (denoted *RF-Hier-MC*) in terms of average sensitivity and F-measure. However, as shown in Fig. 4, the performance of the two classifiers varies significantly across CKD stages.

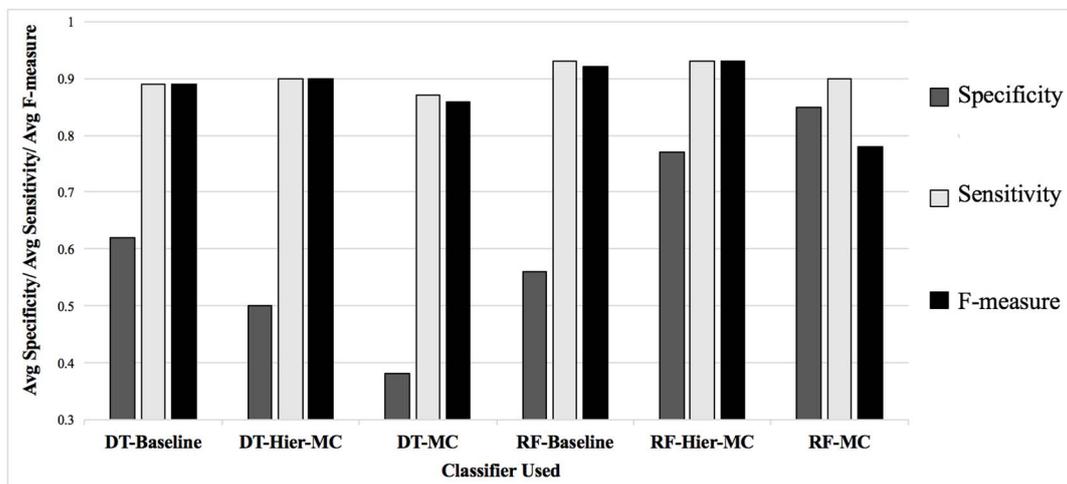

Figure 3. Average specificity, average sensitivity and average F-measure of the decision tree (*DT*) and random forest (*RF*) baseline classifier (*Baseline*), meta-classifier (*MC*) and hierarchical meta-classifier (*Hier-MC*) for assigning CKD severity stages (*3-5*) to patient's office visit records. Classifiers were trained on office visit records from *2007-14* using the complete features set (*490* features) to represent patients, while records from *2015* were used as the test set. The X-axis shows the classifier used; the Y-axis shows the average specificity, sensitivity and F-measure for each of the classifiers.

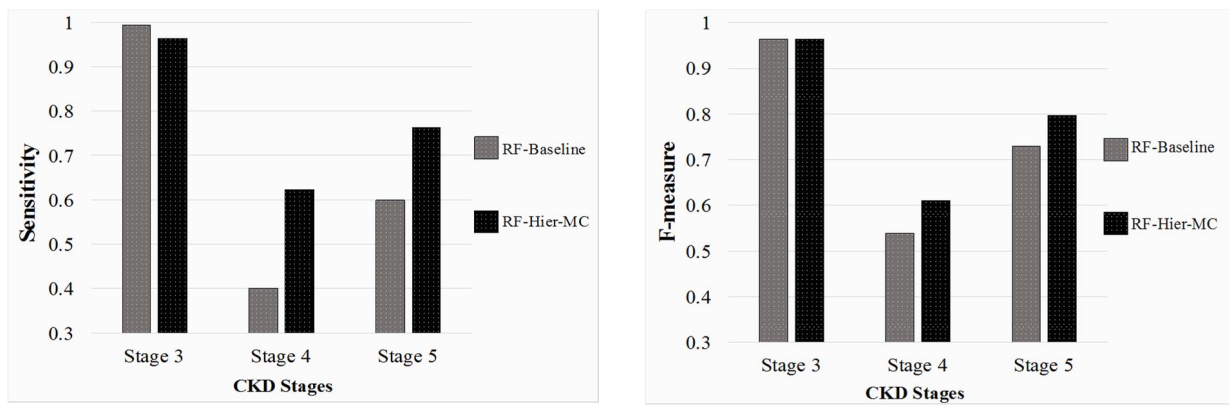

Figure 4. Performance per-class in terms of Sensitivity (left plot) and F-measure (right plot), of the random forest baseline classifier (*RF-Baseline*) and of the random forest hierarchical meta-classifier (*RF-Hier-MC*). The X-axes of both plots show CKD stages; the Y-axes show sensitivity and F-measure per CKD stage.

Clearly, the hierarchical meta-classifier shows a much higher sensitivity and F-measure for both stages *4* and *5* than the baseline classifier, indicating that the hierarchical meta-classifier identifies *advanced* CKD stage records (stages *4* and *5*) more effectively than the baseline. Fig. 3 and Fig. 4 both demonstrate that hierarchical meta-classification outperforms the baseline and the simple meta-classifier for identifying CKD stages *4* and *5*.

We note that in the context of risk-stratification, and particularly when assigning an advanced stage label to a record, *false negatives* (i.e. missing a severe case) have much more severe implications than *false positives* (assigning a stage *4* or *5* label to a stage *3* case). That said, having a very large portion of false-positives is clearly undesirable as it generates false alarms. We further examine these points by calculating the *precision*, (also referred to as *positive predictive value, PPV*) and the *sensitivity* for the set of records associated with the combined stage *4* and *5* records. Precision penalizes for false positives, while sensitivity penalizes for false negatives.

The precision of the classification with respect to the combined set is *0.68*. That is, of the *1,157* records classified as stages *4* or *5* by the hierarchical meta-classifier in our test set, *787* are correctly identified. It is important to note that of the remaining *370* false-positive records, *181* (*~50%*) are *borderline cases,* as indicated by eGFR values in the range of *30-44*, which is associated with patients suffering from advanced stage *3* CKD (stage *3b*) [21]. Recent studies demonstrate that stage *3b* is the inflection point for adverse outcomes, including progression to end stage renal disease (stage *5*) [22]. Thus, the random forest hierarchical meta-classifier effectively identifies not only the advanced stage records already marked, but also the likely-to-be severe cases that are not yet labeled as such. As for the sensitivity, Fig. 4 clearly shows that hierarchical meta-classifier has a much higher sensitivity for stages *4* and *5* records than the baseline, while retaining about the same sensitivity as the baseline with respect to stage *3*.

To verify that the features detected as pertinent by our feature selection method are indeed known to be predictive of CKD severity-levels, we conducted a survey of the medical literature. The survey suggests that most of the features identified by *LASSO* regression are known to be indicative of kidney disease, providing validation that our feature selection identifies attributes that are expected to be predictive of CKD severity-levels [2,22]. For instance, conditions such as *Congestive Heart Failure, Type 2 Diabetes Mellitus* and *Anemia*, which the method identifies as informative, are widely known to be associated with kidney disease [2, 22].

Our results from the experiments in which only the *119* selected features were used, (Tables 3,4) demonstrate that even for the reduced feature set, our method shows high average accuracy, sensitivity, precision and F-measure values of *0.92*. Notably, the difference between the results obtained when using *119* features and those obtained when using the complete set of *490* features, according to all five measures, is very small. The consistent high performance values of our model when trained on a reduced feature set, indicates that the identified features are indeed highly informative of CKD.

Moreover, the high performance values obtained from the random forest hierarchical meta-classifier when trained on the datasets obtained by progressively truncating the early years of patient history, one year at time, indicate that the model effectively identifies CKD stages even when it is trained on limited, recent patient history. As can be seen from Fig. 5, the true-positive rate remains constant, regardless of the number of years included in the visit record, except for a slight decline in predicting stages *4* and *5* when using data from *2013/14* or *2014* alone. The false-negative rate is not impacted by the reduction in a patient's history.

Our results thus indicate that training our model on limited patient history does not significantly affect performance. Similar results were also obtained when the number of features was reduced. The consistent good performance of our method for datasets of different sizes containing patients' visit records gathered over different year ranges, indicates our model's stability and generalizability, and highlights its applicability in clinical settings, where old records are not always available to train the model.

TABLE 3. EXAMPLES OF *18* OF THE *119* FEATURES IDENTIFIED AS INFORMATIVE OF CKD STAGES BY OUR FEATURE SELECTION METHOD, *LASSO* REGRESSION.

| Category | Example Features |
|---|---|
| **Demographics** | *Age*; *Gender*; *Race* |
| **Vital Signs** | *Body Mass Index*; *Systolic Blood Pressure*; *Diastolic Blood Pressure* |
| **Medications** | *Insulin Response Enhancers Biguanides*; *Direct Acting Vasodilators*; *Insulin*; *ACE Inhibitors*; *Beta Blockers Cardiac Selective*; *Alpha Beta Blockers*; *Antihyperglycemic Dipeptidyl Peptidase 4 Inhibitors* |
| **Diagnosed Conditions** | *Congestive Heart Failure*; *Type 2 Diabetes Mellitus*; *Anemia*; *Erythrocyte sedimentation rate raised*; *Benign Essential Hypertension*; *Degenerative joint disease of pelvis* |

TABLE 4. COMPARISON OF RANDOM FOREST HIERARCHICAL META-CLASSIFIER (*RF-HIER-MC*) TRAINED ON DATASETS REPRESENTED VIA *119* FEATURES AND A CLASSIFIER TRAINED ON A DATASET REPRESENTED VIA *490* FEATURES. STANDARD DEVIATION IS SHOWN IN PARENTHESES.

| *Classifier* | Accuracy | Specificity | Sensitivity | Precision | F-measure |
|---|---|---|---|---|---|
| **RF-Hier-MC (*119 features*)** | 0.92 (*0.02*) | 0.74 (*0.03*) | 0.92 (*0.02*) | 0.92 (*0.02*) | 0.92 (*0.02*) |
| **RF-Hier-MC (*490 features*)** | 0.93 (*0.02*) | 0.77 (*0.02*) | 0.93 (*0.02*) | 0.93 (*0.02*) | 0.93 (*0.02*) |

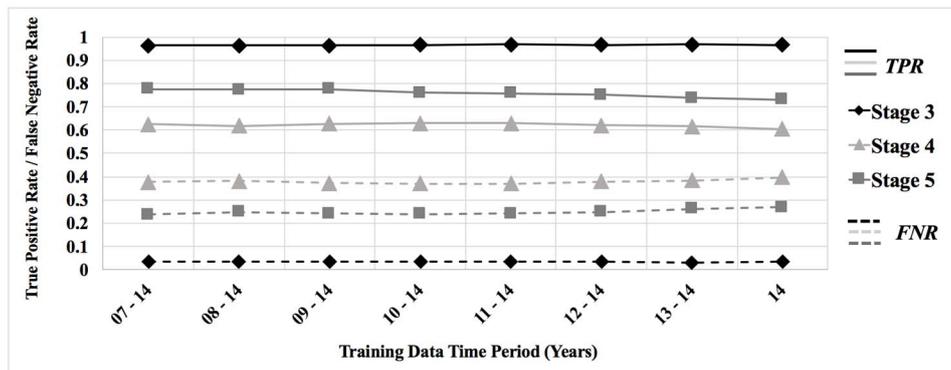

Figure 5. True Positive Rates, *TPR* (*solid plots*) and False Negative Rates, *FNR* (*dashed plots*) with respect to stages *3*, *4* and *5*, associated with eight hierarchical meta-classifiers, each trained on datasets obtained by gradually truncating the early years of patient history included in the training set, one year at a time. The X-axis shows the years covered by each training set. The Y-axis shows the true positive rate (top) or false negative rate (bottom).

## VII. CONCLUSION

We have shown, using a large collection of office visit records gathered from many thousands of patients, that CKD severity-levels can be effectively stratified using simple quantitative attributes collected during standard office visits, in the face of imbalanced data. We developed a hierarchical meta-classifier to assess CKD stages from highly imbalanced training sets, achieving average accuracy, sensitivity, precision and F-measure of *0.93*. Our method significantly outperforms baseline classifiers and simple meta-classifiers in identifying advanced CKD stages (stages *4* and *5*). Our results also show that the method retains its good performance when the feature set is reduced, as well as when the number of records is significantly truncated.

As a future direction, we plan to evaluate the efficacy of the method in separating patient records not associated with kidney disease or those associated with less-severe stages, from those associated with severe stages. We also expect that our method can be applied to attain severity stratification in other health conditions using standard office visit records.

## ACKNOWLEDGMENT

This research was partially supported by NIGMS IDeA grants U54-GM104941 and P20 GM103446, and by NSF IIS EAGER grant #1650851. We thank James T. Laughery and Sarahfaye Dolman for their major role in building the dataset.